\documentclass[12pt]{article}
\usepackage{amssymb,amsmath,epsfig}

\begin{document}

\title{\bf Possible Formation of Compact Stars in $f(R,T)$ Gravity}

\author{M. Zubair \thanks{mzubairkk@gmail.com; drmzubair@ciitlahore.edu.pk} ${}^{(a)}$,
 G. Abbas${}^{(b)}$ \thanks{ghulamabbas@ciitsahiwal.edu.pk}
Ifra Noureen \thanks{ifra.noureen@gmail.com} ${}^{(c)}$, \\
${}^{(a)}$ Department of Mathematics,\\ COMSATS Institute of
Information Technology, Lahore, Pakistan.\\ ${}^{(b)}$  Department
of Mathematics,\\ COMSATS Institute of Information Technology,
Sahiwal, Pakistan.
\\ ${}^{(c)}$ Department of
Mathematics,\\University of Management and Technology, Lahore,
Pakistan.}

\date{}
\maketitle
\begin{abstract}

This paper constitutes the investigations regarding possible
formation of compact stars in $f(R,T)$ theory of gravity, where $R$
is Ricci scalar and $T$ is trace of energy momentum tensor. In this
connection, we use analytic solution of Krori and Barua metric
(Krori and Barua 1975)
 to spherically symmetric anisotropic star in context of
$f(R,T)$ gravity. The masses and radii of compact stars models
namely model $1$, model$ 2$ and model $3$ are employed to
incorporate with unknown constants in Krori and Barua metric. The
physical features such as regularity at center, anisotropy measure,
casuality and well-behaved condition of above mentioned class of
compact starts are analyzed. Moreover, we have also discussed energy
conditions, stability and surface redshift in $f(R,T)$ gravity.
\end{abstract}

\section{Introduction}

The late time evolution of stars influenced by strong gravitational
pull has been a largely anticipated field in astrophysics and
gravitational theories. It facilitates in investigations of diverse
characteristics of gravitating sources via various physical
phenomenons. Baade and Zwicky (1934) proposed the inception of
massive compact stellar objects, establishing the argument that
supernova may result in a small and super dense star. This
eventually came in to reality in 1967, when Bell and Hewish (Longair
1994, Ghosh 2007) discovered pulsars that are highly magnetized and
rotating neutron stars. So, in reality, we come across a fundamental
revolutionary shift from normal stars to compact stars, with a wide
range in the form of stars to neutron stars, quarks, dark stars,
gravastars and finally black holes.

In our present work, we are more specifically interested in study of
compact stars category. Generally the homogeneity of the spherically
symmetric matter configuration is emphasized while theoretical
modeling of a compact star, satisfying the Tolman
Oppenheimer-Volkoff (TOV) equation. Ruderman (1972) was the first
one to argue that the nuclear matter density becomes anisotropic at
the core of compact object. Analytic solutions of the field
equations for various static spherically symmetric configurations
for anisotropic compatible to interior of compact stellar modeling
have been obtained in numerous works (Maurya and Gupta 2012,2013
2014, Maharaj et al.2014, Pant et al. 2014a, 2014b). The pressure
inside the fluid sphere disintegrates in two parts, namely radial
and tangential pressure. It has been investigated that anisotropy
gave rise to the repulsive force that assists to construct the
compact objects. In this context, it is established in (Kamal et
al.2012) that the Krori and Barua (henceforth KB)(1975) metric
provides an effective and realistic approach in modeling of compact
stars.

The numerical simulations can be taken into account to explore the
characteristics of compact stars from integrated TOV equations, if
equation of state (EOS) is known. Rahaman et al. (2012) extended the
KB models by using the Chaplygin gas EoS and discussed their
physical features. Mak and Harko (2004) established standard models
of spherically symmetric compact stars via exact solution of the
field equations. They determined the impressions of physical
parameters such as energy density, tangential and radial pressure,
concluding that these parameters remain finite and positive inside
the stars. The anisotropic exact models for compact objects with a
barotropic EOS are discussed in (2006). Hossein et al.(2012) studied
the impact of cosmological constant on anisotropic compact stars.

General Relativity (GR) being fundamental theory of gravity is
successful in weak field limit, but insufficient to explore the
strong field. The expected description of GR in strong field regime
can be done by its modifications. The modifications in
Einstein-Hilbert (EH) action are induced to arrive at alternative
theories of gravity. Among modified theories of gravity $f(R)$
gravity being most elementary modification of GR is extensively
studied in context of existence and stability of neutron stars and
compact stars (Arapoglu et al.2012, Alavirad and Weller 2013,
Astashenok et al. 2014, 2015, Yazadjiev et al. 2014, Kausar and
Noureen 2014, Noureen et al.2015). Abbas and his collaborators
(2014, 2015a, 2015b, 2015c) analyzed a class compact stars in GR,
$f(T)$,(where $T$ is torsion scalar) with different equation of
state.

The issue of accelerated expansion of the universe can be explained
by taking into account the modified theories of gravity such as
$f(R,T)$ gravity (Harko et al.2011). The $f(R,T)$ gravity provides
an alternative way to explain the current cosmic acceleration with
no need of introducing either the existence of extra spatial
dimension or an exotic component of dark energy. Harko et al.(2011)
generalized $f(R)$ gravity by introducing an arbitrary function of
the Ricci scalar $R$ and the trace of the energy-momentum tensor
$T$. The dependence of $T$ may be introduced by exotic imperfect
fluids or quantum effects (conformal anomaly). As a result of
coupling between matter and geometry, motion of test particles is
nongeodesic and an extra acceleration is always present. In $f(R,T)$
gravity, cosmic acceleration may result not only due to geometrical
contribution to the total cosmic energy density but it also depends
on matter contents.

 Soon after the origination of $f(R,T)$ gravity,
its cosmological and thermodynamic implications including the energy
conditions and dynamical analysis were extensively discussed
(Shabani and Farhoudi 2014, Harko and Lobo 2010, Harko 2010, Azizi
2013, Sharif and Zubair 2012a, 2012b, 2013, Jamil et al. 2012a,
2012b, Momeni et al. 2015a, 2015b, Momeni and Myrzakulov 2015,
Barrientos and Rubilar 2014). However, the explorations regarding
compact stars in $f(R,T)$ gravity are yet to be done. Herein, we are
interested to study the structure of a class of compact stars Model
1, Model 2 and Model 3 in $f(R,T)$ gravity.

The modified action in $f(R,T)$ is as follows(Harko et al. 2011)
\begin{equation}\label{1}
\int dx^4\sqrt{-g}[\frac{f(R, T)}{16\pi G}+\mathcal{L} _ {(m)}],
\end{equation}
where $\mathcal{L} _ {(m)}$ is matter Lagrangian and $g$ denote the
metric tensor. Different choices of $\mathcal{L} _ {(m)}$ can be
considered, each of which directs to a specific form of fluid. In
our present work the viable $f(R, T)$ model we have chosen is of
following type
\begin{equation}
\label{2} f(R, T)=f_1(R) +f_2(T),
\end{equation}
where $f_1(R)$ is a function of Ricci scalar and $\lambda$ is some
positive constant value. We have analyzed above mentioned $f(R, T)$
model with $f_1(R)=R+\alpha{R}^2$, $\alpha$ being a positive scalar.
In our discussion, we have considered the model $f(R,T)=f_1(R)
+f_2(T)$ which does not imply the direct non-minimal gravitational
coupling between scalar curvature $R$ and trace of the
energy-momentum tensor $T$ in the Lagrangian level. But there may be
coupling between matter and geometry which becomes apparent in the
study of thermodynamics (Sharif and Zubair 2012a). The  $f_2(T)$ can
be considered as matter correction term to $f(R)$ gravity and in
this particular study we choose $f_2(T)={\lambda}T$. The main reason
behind the difference on cosmology in ordinary $f(R)$ gravity and in
the above $f(R,T)$ model is the non-trivial coupling between matter
and geometry. In our previous work (Sharif and Zubair 2012b), we
have reconstructed some explicit models of this $f(R,T)$ gravity for
anisotropic universe and explored the phantom era of dark energy.

We have investigated that only for $\lambda=1$ and $\alpha=2$, the
energy density of all the considered models remains positive and
energy conditions are only valid for these values of $\lambda$ and
$\alpha$. So, through out the analysis, we have used the above
values of $\lambda$ and $\alpha$, it will not be mentioned again
explicitly.

This paper is arranged as: In the following section anisotropic
matter distribution and expressions physical parameters for energy
density $\rho$, radial and tangential pressure are established.
Section \textbf{3} constitutes the analysis of physical features of
compact stars and their stability analysis. In the last section, we
have summarized our results.

\section{Anisotropic Matter Distribution in $f(R, T)$ Gravity}

The line element for particular spherically symmetric metric
describing the compact star stellar configuration is
\begin{equation}\label{6}
ds^2=e^{a(r)}dt^{2}-e^{b(r)}dr^{2}-r^2(d\theta^{2}+\sin^{2}\theta
d\phi^{2}),
\end{equation}
where $b=Ar^2$, $a=Br^2+C$ (Krori and Barua 1975), $A$, $B$ and $C$ are arbitrary
constant that will be calculated by using some physical assumptions.
The above set of functions are introduced to arrive at singularity
free structure for compact star. Clearly, this set of functions
leads to non singular density and curvature setting.

Taking $8\pi G = 1$ and upon variation of modified EH action in
$f(R,T)$ (\ref{1}) with respect to metric tensor $g_{uv}$, the
following modified field equations are formed as
\begin{eqnarray}\nonumber
G_{uv}&=&\frac{1}{f_R}\left[(f_T+1)T^{(m)}_{uv}-\rho g_{uv}f_T+
\frac{f-Rf_R}{2}g_{uv}\right.\\\label{3}&+&\left.(\nabla_u\nabla_v-g_{uv}\Box)f_R\right],
\end{eqnarray}
where $f_R=\frac{\partial f(R,T)}{\partial R}$ , $f_T=\frac{\partial
f(R,T)}{\partial T}$  and  $T^{(m)}_{uv}$ denotes the usual matter
energy momentum tensor that is considered to be anisotropic, is
given by
\begin{equation}\label{4}
T^{(m)}_{uv}=(\rho+p_{t})V_{u}V_{v}-p_{t}g_{uv}+(p_{r}-p_{t})\chi_{u}\chi_{v},
\end{equation}
where $\rho$, $p_r$ and $p_t$ denote energy density, radial and
transverse stresses respectively. The four velocity is denoted by
$V_{u}$ and $\chi_{u}$ to be the radial four vector satisfying
\begin{equation}\label{5}
V^{u}=e^{\frac{-a}{2}}\delta^{u}_{0},\quad V^{u}V_{u}=1,\quad
\chi^{u}=e^{\frac{-b}{2}}\delta^u_1,\quad \chi^{u}\chi_{u}=-1.
\end{equation}

When $f(R, T)=f_1(R) +\lambda T$, the expression for $\rho$, $p_r$
and $p_t$ can be extracted from modified field equations as follows

\begin{eqnarray}\nonumber
\rho&=&
\frac{1}{2(1+2\lambda)}\left[\frac{2+5\lambda}{e^b(1+\lambda)}
\left\{\left(\frac{a'}{r}-\frac{a'b'}{4}+\frac{a''}{2}
+\frac{a'^2}{4}\right)f_{1R}-
f''_{1R}+\left(\frac{b'}{2}-\frac{f_1}{2}e^b\right.\right.\right.\\\nonumber&&\left.\left.\left.
-\frac{2}{r}\right)f'_{1R}\right\}+
\frac{\lambda}{e^b(1+\lambda)}\left\{\left(\frac{a'b'}{4}+\frac{b'}{r}-\frac{a''}{2}
-\frac{a'^2}{4}\right)f_{1R}
+\left(\frac{a'}{2}+\frac{2}{r}\right)f'_{1R}\right.\right.\\\nonumber&&\left.\left.+\frac{f_1}{2}e^b\right\}+
\frac{2\lambda}{e^b(1+\lambda)}\left\{\frac{f_{1R}}{r^2}\left(\frac{(b'-a')r}{2}-e^b+1\right)
+\left(\frac{a'-b'}{2}+\frac{1}{r}\right)f'_{1R}\right.\right.\\\label{7}&&\left.\left.+f''_{1R}
+\frac{f_1}{2}e^b\right\}\right],
\\\nonumber
p_r&=& \frac{-1}{2(1+2\lambda)}\left[\frac{\lambda}{e^b(1+\lambda)}
\left\{\left(\frac{a'}{r}-\frac{a'b'}{4}+\frac{a''}{2}
+\frac{a'^2}{4}\right)f_{1R}-
f''_{1R}+\left(\frac{b'}{2}-\frac{f_1}{2}e^b\right.\right.\right.\\\nonumber&&\left.\left.\left.
-\frac{2}{r}\right)f'_{1R}\right\}-
\frac{(2+3\lambda)}{e^b(1+\lambda)}\left\{\left(\frac{a'b'}{4}+\frac{b'}{r}-\frac{a''}{2}
-\frac{a'^2}{4}\right)f_{1R}
+\left(\frac{a'}{2}+\frac{2}{r}\right)f'_{1R}\right.\right.\\\nonumber&&\left.\left.+\frac{f_1}{2}e^b\right\}+
\frac{2\lambda}{e^b(1+\lambda)}\left\{\frac{f_{1R}}{r^2}\left(\frac{(b'-a')r}{2}-e^b+1\right)
+\left(\frac{a'-b'}{2}+\frac{1}{r}\right)f'_{1R}\right.\right.\\\label{8}&&\left.\left.+f''_{1R}
+\frac{f_1}{2}e^b\right\}\right],
\\\nonumber
p_t&=& \frac{-1}{2(1+2\lambda)}\left[\frac{\lambda}{e^b(1+\lambda)}
\left\{\left(\frac{a'}{r}-\frac{a'b'}{4}+\frac{a''}{2}
+\frac{a'^2}{4}\right)f_{1R}-
f''_{1R}+\left(\frac{b'}{2}-\frac{f_1}{2}e^b\right.\right.\right.\\\nonumber&&\left.\left.\left.
-\frac{2}{r}\right)f'_{1R}\right\}-
\frac{\lambda}{e^b(1+\lambda)}\left\{\left(\frac{a'b'}{4}+\frac{b'}{r}-\frac{a''}{2}
-\frac{a'^2}{4}\right)f_{1R}
+\left(\frac{a'}{2}+\frac{2}{r}\right)f'_{1R}\right.\right.\\\nonumber&&\left.\left.
+\frac{f_1}{2}e^b\right\}-
\frac{2}{e^b}\left\{\frac{f_{1R}}{r^2}\left(\frac{(b'-a')r}{2}-e^b+1\right)
+\left(\frac{a'-b'}{2}+\frac{1}{r}\right)f'_{1R}\right.\right.\\\label{9}&&\left.\left.+f''_{1R}
+\frac{f_1}{2}e^b\right\}\right].
\end{eqnarray}
Here $f_{1R}=\frac{df_1}{dR}$ and prime denotes the derivatives with
respect to radial coordinate. Substituting Eq.(\ref{2}) in
Eqs.(\ref{7})-(\ref{9}) and inserting value of Ricci scalar in the
form of metric functions together with the KB metric coefficients,
we arrive at
\begin{eqnarray}\nonumber
\rho&=&\frac{1}{r^4(1+3\lambda+2\lambda^2)}\left[e^{-2Ar^2}\left\{e^{2Ar^2}(r^2+2\alpha
(\lambda-1))+2\alpha(-1-3Br^2\right.\right.\\\nonumber&&\left.\left.
-B^2r^4+Ar^2(2+Br^2))(1-\lambda-3Br^2(1+3\lambda)
+B^2r^4(3+7\lambda))\right.\right.\\\nonumber&&\left.\left.
-e^{Ar^2}(4\alpha(\lambda-1)+r^2(1+24B\alpha\lambda)+8A
\alpha(1+\lambda)+Br^6(A\lambda+B(2\right.\right.\\\label{10}&&\left.\left.
+4\lambda)))-r^4(A(2+(4+8B\alpha)\lambda)+B(3\lambda+
4B\alpha(2+3\lambda)))\right\}\right],
\\\nonumber
p_r&=&\frac{1}{r^4(1+3\lambda+2\lambda^2)}\left[e^{-2Ar^2}\left\{e^{2Ar^2}(r^2+2\alpha
(\lambda-1)+e^{Ar^2}(4\alpha(\lambda-1)\right.\right.\\\nonumber&&\left.\left.
+ABr^6\lambda+r^2(1+8B\alpha
(\lambda-1)-8A\alpha\lambda)+Br^4(2+\lambda-4\alpha\lambda(2A-B)))
\right.\right.\\\nonumber&&\left.\left.-
2\alpha(1+3Br^2+B^3r^4-Ar^2(2+Br^2))(-1+Br^2(\lambda-1)-Ar^2(2(1
\right.\right.\\\label{11}&&\left.\left.+\lambda)
+Br^2(1+3\lambda)))\right\}\right],
\\\nonumber
p_t&=&\frac{1}{r^4(1+3\lambda+2\lambda^2)}\left[e^{-2Ar^2}\left\{2\alpha(1+3Br^2+B^2r^4
-Ar^2(2+Br^2))(-3\right.\right.\\\nonumber&&\left.\left.
-B^2r^4-2\lambda+Br^2(Ar^2-1)(1+2\lambda))+e^{2Ar^2}(r^2(2+\lambda)-2\alpha
(3+2\lambda))\right.\right.\\\nonumber&&\left.\left.
+e^{Ar^2}(4\alpha(3+2\lambda)+Br^6(A-B+A\lambda)-r^2(2+\lambda+4A\alpha(3+2\lambda)
\right.\right.\\\nonumber&&\left.\left.-4B\alpha(5+4\lambda))+r^4(B(4B\alpha-1)
(2+\lambda)+A(1-8B\alpha(1+\lambda))))
\right\}\right].\\\label{12}
\end{eqnarray}

The Schwarzschild solution is considered to be most suitable choice
for matching conditions in exterior regime (Noureen and Zubair 2015,
2014, Goswami et al.2014 Cooney et al.2010, Ganguly et al. 2014).
The interior metric of the boundary surface will be the same for the
internal or external geometry of the star. Herein, the exterior
metric of the star is described by the Schwarzschild solution, given
by
 \begin{equation}\label{21}
 ds^2=\left(1-\frac{2M}{r}\right)dt^2-\left(1-
 \frac{2M}{r}\right)^{-1}dr^2-r^2(d\theta^2+ sin^2{\theta}d\varphi^2),
\end{equation}
Smooth matching of the interior metric (\ref{6}) to the vacuum exterior solution at the boundary surface $r=R$, yield
\begin{eqnarray}\label{22}
  g_{tt}^-=g_{tt}^+,~~~~~
   g_{rr}^-=g_{rr}^+,~~~~~
   \frac{\partial g_{tt}^-}{\partial r}=\frac{\partial g_{tt}^+}{\partial r},
  \end{eqnarray}
where superscript $-$ and $+$ stands for interior and exterior solutions.
Matching of the interior and exterior spacetime leads to following
 \begin{eqnarray}\label{23}
  A&=&-\frac{1}{R^2}ln\left(1-\frac{2M}{R}\right),\\\label{24}
 B&=&\frac{M}{R^3}{{\left(1-\frac{2M}{R}\right)}^{-1}},\\\label{24a}
 C&=&ln\left(1-\frac{2M}{R}\right)-\frac{M}{R}{{\left(1-\frac{2M}{R}\right)}^{-1}}.
\end{eqnarray}
The values of the constants $A$ and $B$ are evaluated by using
approximate values of $M$ and $R$ (Lattimer and Steiner 2014, Li et
al.1999) of the considered compact stars, provided in the table
\textbf{1} . The compactness of a star can be defined by
$u=\frac{M(R)}{R}$ and surface redshift $Z_s$ can be determined by
using the result $Z _s=(1-2u)^{-1/2}-1$. The values of $Z_s$ for the
considered models have been given in table \textbf{1}.
\begin{table}[ht]
\caption{Approximate Values of the model parameters for considered
compact stars}
\begin{center}
\begin{tabular}{|c|c|c|c|c|c|c|}
\hline {Models}&  \textbf{ $M$} & \textbf{$R(km)$} & \textbf{
$u=\frac{M}{R}$} &\textbf{ $A(km ^{-2})$}& \textbf{$B(km ^{-2})$}&
\textbf{$Z_s$}
\\\hline  Model 1& 0.88$M_\odot$& 7.7&0.168&0.006906276428 &0.004267364618 & $0.23$
\\\hline Model 2& 1.435$M_\odot$& 7.07&0.299& 0.01823156974 &0.01488011569 & $0.57$
\\\hline Model 3&2.25$M_\odot$& 10.0 &0.332&0.01090644119 &0.009880952381 & $0.0.73$
\\\hline
\end{tabular}
\end{center}
\end{table}

\section{Physical Analysis}

This section covers the physical constraints required for interior
solution, incorporating anisotropic behavior, matching and energy
conditions together with the stability analysis of considered
compact stars.

\subsection{Anisotropic Behavior}

Prior to the discussion of anisotropy measure, we discuss the evolution of
energy density $\rho$ and anisotropic stresses $p_r$ and $p_t$ respectively, shown
in Figures \textbf{1-3}.
\begin{figure}
\centering \epsfig{file=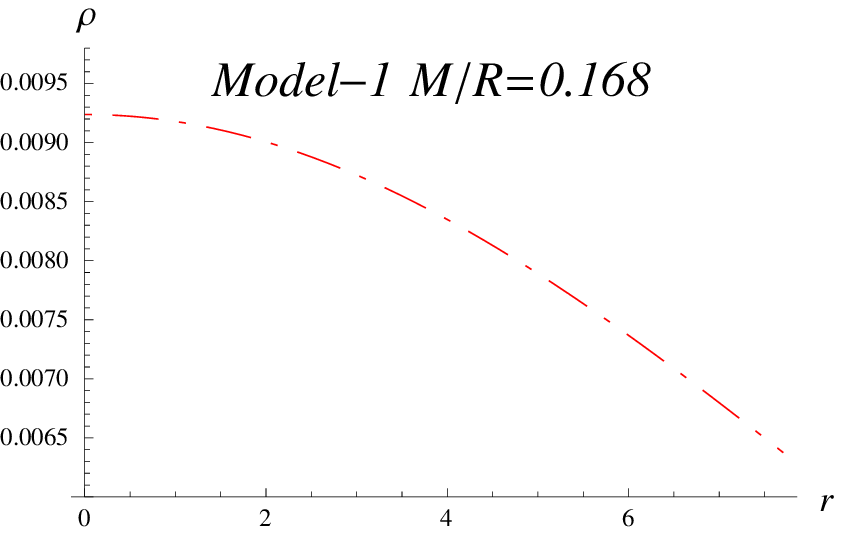, width=.34\linewidth,
height=1.4in}\epsfig{file=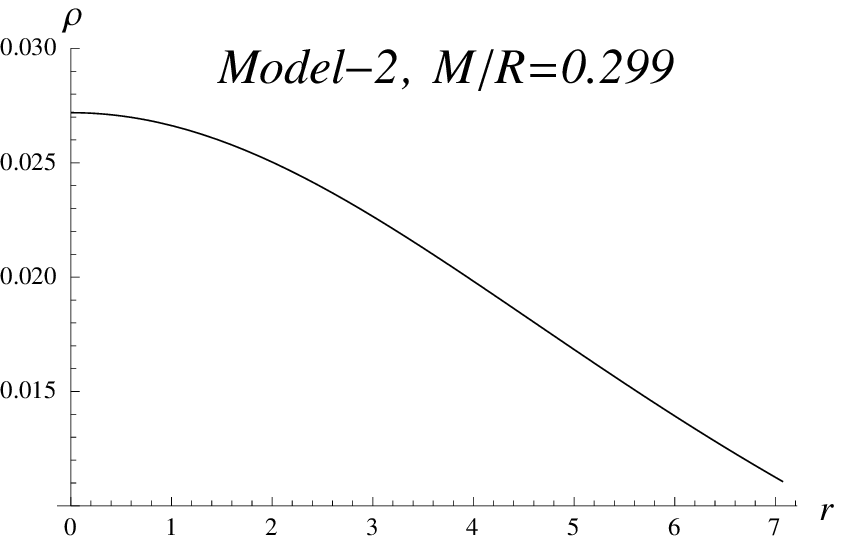, width=.36\linewidth,
height=1.4in}\epsfig{file=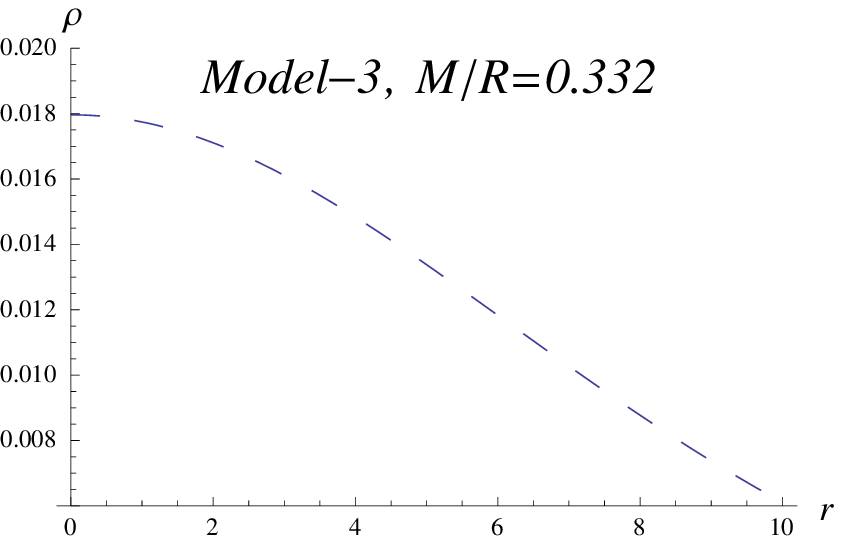, width=.34\linewidth,
height=1.4in}\caption{Evolution of energy density $\rho$ with radius $r$}
\end{figure}
\begin{figure}
\centering \epsfig{file=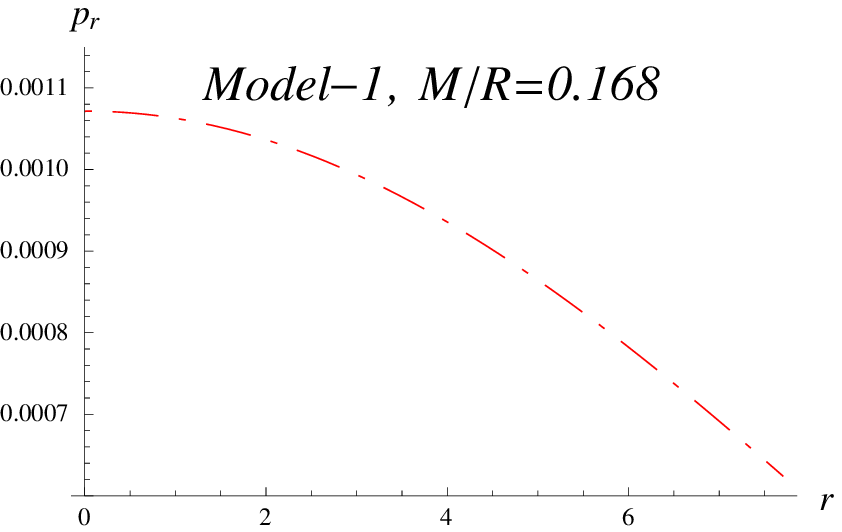, width=.34\linewidth,
height=1.4in}\epsfig{file=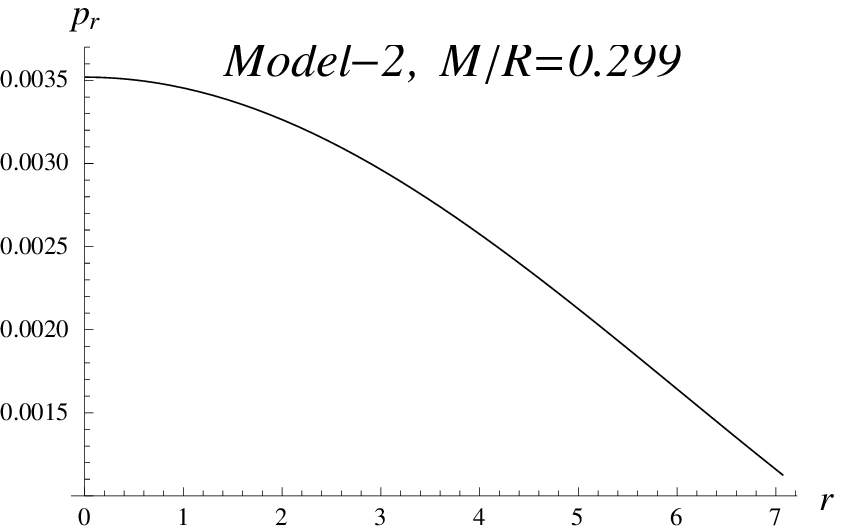, width=.36\linewidth,
height=1.4in}\epsfig{file=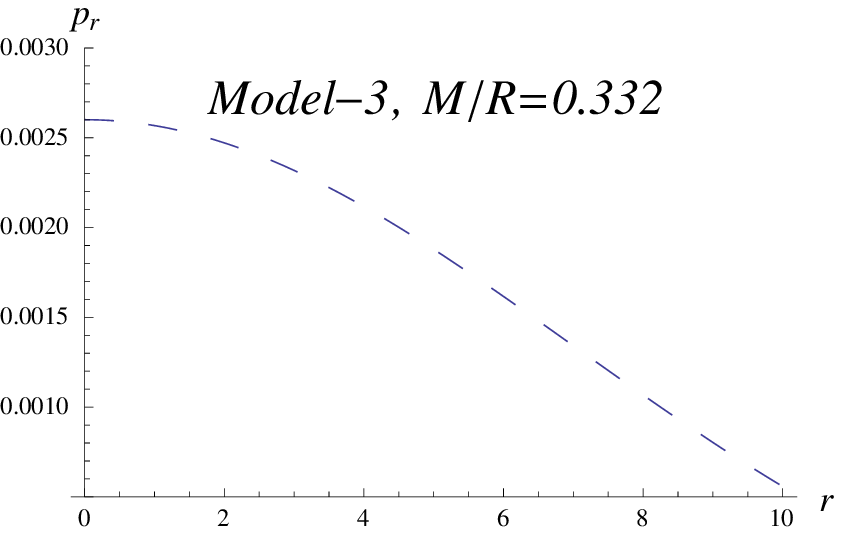, width=.34\linewidth,
height=1.4in}\caption{Variation of radial pressure $p_r$ with radius $r$}
\end{figure}
\begin{figure}
\centering \epsfig{file=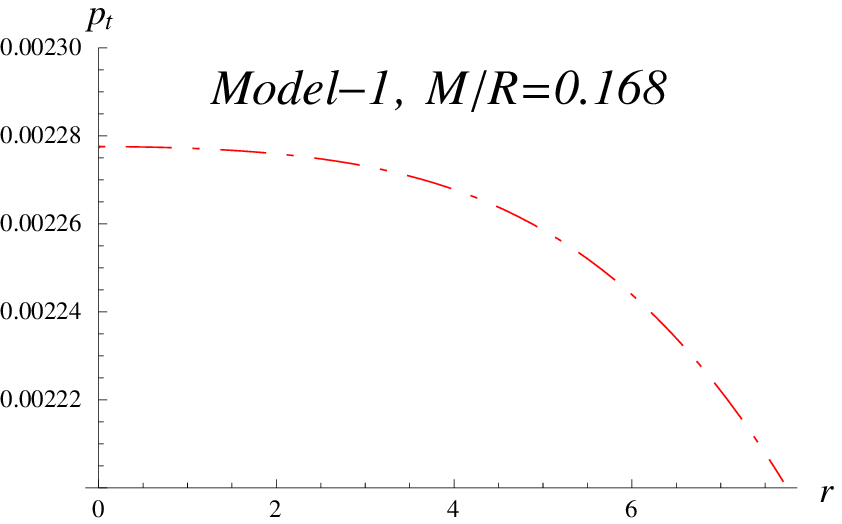, width=.34\linewidth,
height=1.4in}\epsfig{file=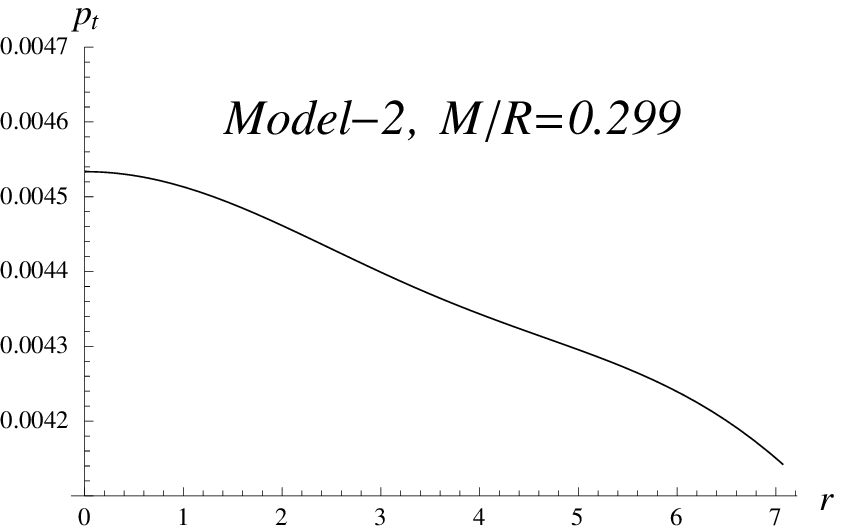, width=.36\linewidth,
height=1.4in}\epsfig{file=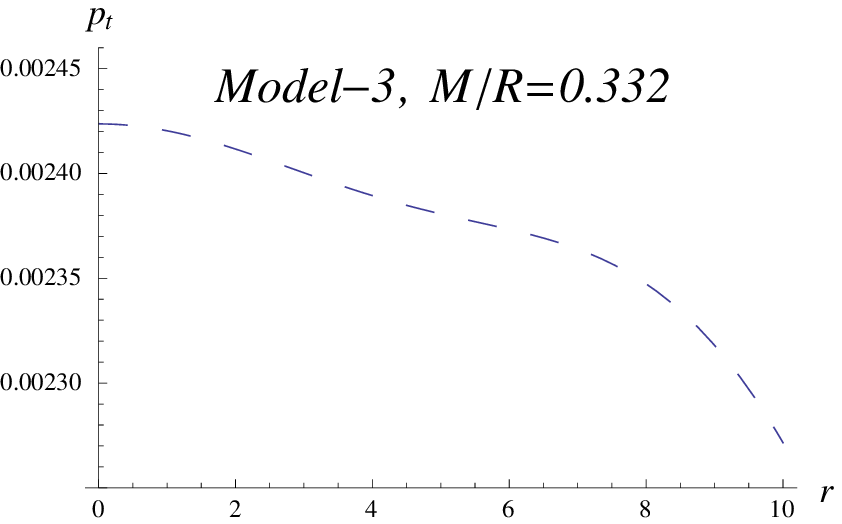, width=.34\linewidth,
height=1.4in}\caption{Evolution of tangential pressure $p_r$ with radius $r$}
\end{figure}
Radial derivative of equation (\ref{7}) leads to the following expression
\begin{eqnarray}\nonumber
\frac{d\rho}{dr}&=&\frac{-4 e^{-2 A r^2}}{r^5 (1 + \lambda) (1 + 2 \lambda)}
     (e^{2 A r^2} (r^2 + 2 \alpha (-1+ \lambda)) +
      2 (-1 - 3 B r^2 - B^2 r^4 \\\nonumber&&+A r^2 (2 + B r^2)) \alpha (1 - \lambda -
         3 B r^2 (1 + 3 \lambda) +A r^2 (-2 + B r^2) (1 + 3 \lambda) \\\nonumber&&+
         B^2 r^4 (3 + 7 \lambda)) -e^{ A r^2} (4 \alpha (-1 + \lambda) +
         r^2 (1 + 24 B \alpha \lambda +  8 A \alpha (1 + \lambda)) \\\nonumber&&+
         B r^6 (A \lambda + B (2 + 4 \lambda)) -
         r^4 (A (2 + (4 + 8 B \alpha) \lambda) +
            B (3 \lambda + 4 B \alpha (2 + 3 \lambda)))))\\\nonumber&& -
 \frac{4Ae^{-2 A r^2}}{r^3 (1 + \lambda) (1 + 2 \lambda)} (e^{2 A r^2} (r^2 + 2 \alpha (-1 + \lambda)) +
     2 (-1 - 3 B r^2 - B^2 r^4 + A r^2 (2 \\\nonumber&&+ B r^2)) \alpha(1 - \lambda -
        3 B r^2 (1 + 3 \lambda) +  A r^2 (-2 + B r^2) (1 + 3 \lambda) +
        B^2 r^4 (3 + 7 \lambda))\\\nonumber&& -     e^{ A r^2}(4 \alpha (-1 + \lambda) +
        r^2 (1 + 24 B \alpha \lambda +    8 A \alpha (1 + \lambda)) +
        B r^6 (A \lambda + B (2 + 4 \lambda)) \\\nonumber&&-  r^4 (A (2 + (4 + 8 B \alpha) \lambda) +
           B (3 \lambda + 4 B \alpha (2 + 3 \lambda))))) - \frac{e^{-2 A r^2}}{r^4 (1 + \lambda) (1 + 2 \lambda)}
    \\\nonumber&&\times(2 e^{ 2A r^2} r +  4 A e^{2 A r^2} r (r^2 + 2 \alpha (-1 +\lambda)) +
     2 (-1 - 3 B r^2 - B^2 r^4 +  A r^2 (2\\\nonumber&& + B r^2)) \alpha(-6 B r (1 + 3 \lambda]) +
        2 A B r^3 (1 + 3 \lambda) + 2 A r (-2 + B r^2) (1 + 3 \lambda) +
         (3 \\\nonumber&&+ 7 \lambda)4 B^2 r^3) +
     2 (-6 B r + 2 A B r^3 - 4 B^2 r^3 +        2 A r (2 + B r^2)) \lambda (1 - \lambda -
        3 B r^2 (1 \\\nonumber&&+ 3 \lambda) +    A r^2 (-2 + B r^2) (1 + 3 \lambda) +
        B^2 r^4 (3 + 7 \lambda)) -     e^{A r^2} (2 r (1 + 24 B \alpha \lambda +
            (1 \\\nonumber&&+ \lambda)8 A \lambda) +        6 B r^5 (A \lambda + B (2 + 4 \lambda)) -
        4 r^3 (A (2 + (4 + 8 B \alpha) \lambda) +
           B (3 \lambda + (2 \\\nonumber&&+ 3 \lambda)4 B \alpha ))) -
     2 A  e^{A r^2}
       r (4 \alpha (-1 + \lambda) +
        r^2 (1 + 24 B \alpha \lambda +
           8 A \alpha (1 + \lambda)) +
        B r^6 \\\nonumber&&\times(A \lambda + B (2 + 4 \lambda)) -
        r^4 (A (2 + (4 + 8 B \alpha) \lambda) +
           B (3 \lambda + 4 B \alpha (2 + 3 \lambda)))))<0,\\\label{15}
\end{eqnarray}
and also Eq.(\ref{8}) reveals that $\frac{dp_r}{dr}<0$, depicting decrease in energy density and radial pressure with increasing radius of the compact object, well
supported by the results shown in Figures \textbf{4} and \textbf{5}.
Maximality of central density and pressure is achievable at $r=0$, indicating
\begin{eqnarray}\nonumber&&
\frac{d\rho}{dr}=0, \quad \frac{dp_r}{dr}=0,
\\\nonumber&&
\frac{d^2\rho}{dr^2}<0, \quad \frac{d^2p_r}{dr^2}<0
\end{eqnarray}
\begin{figure}
\centering \epsfig{file=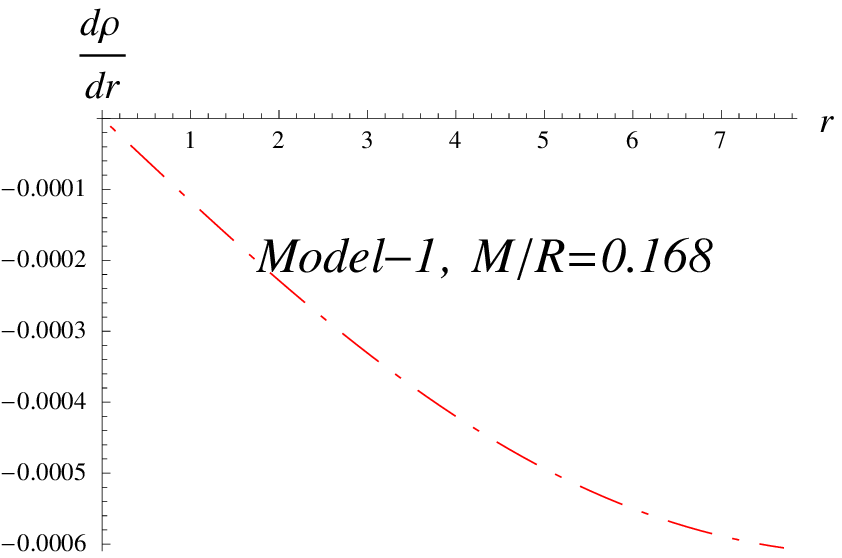, width=.34\linewidth,
height=1.4in}\epsfig{file=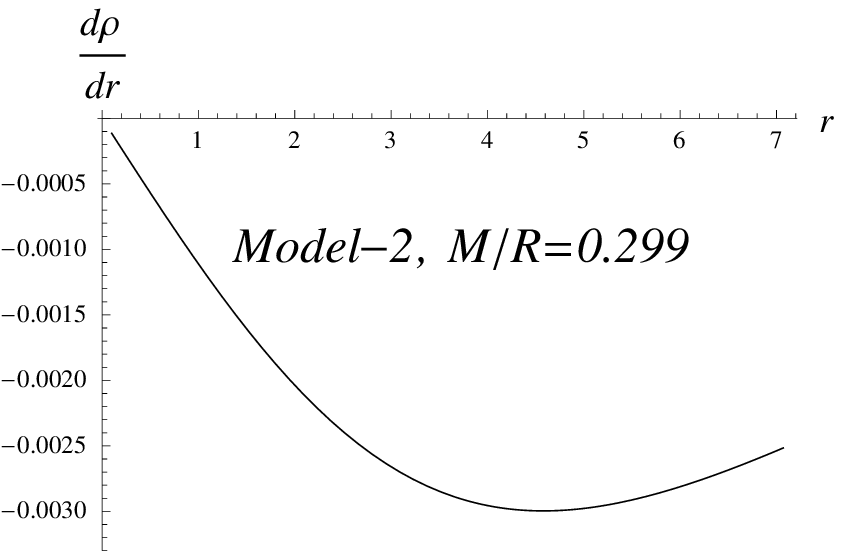, width=.36\linewidth,
height=1.4in}\epsfig{file=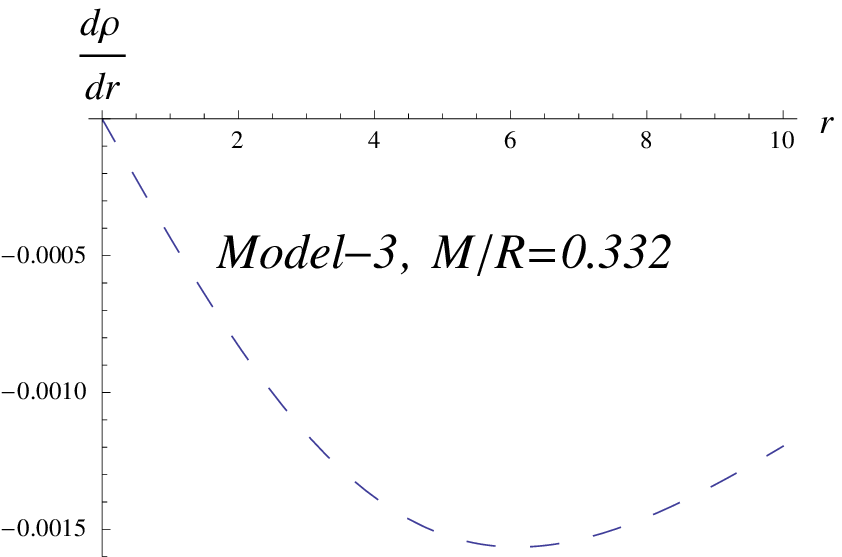, width=.34\linewidth,
height=1.4in}\caption{Behavior of $\frac{d\rho}{dr}$ with increasing radius $r$.}
\end{figure}
\begin{figure}
\centering \epsfig{file=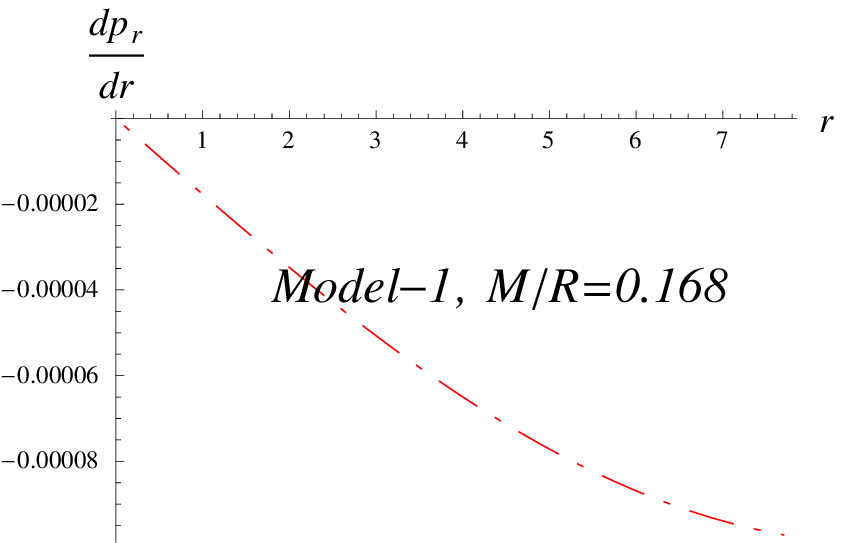, width=.34\linewidth,
height=1.4in}\epsfig{file=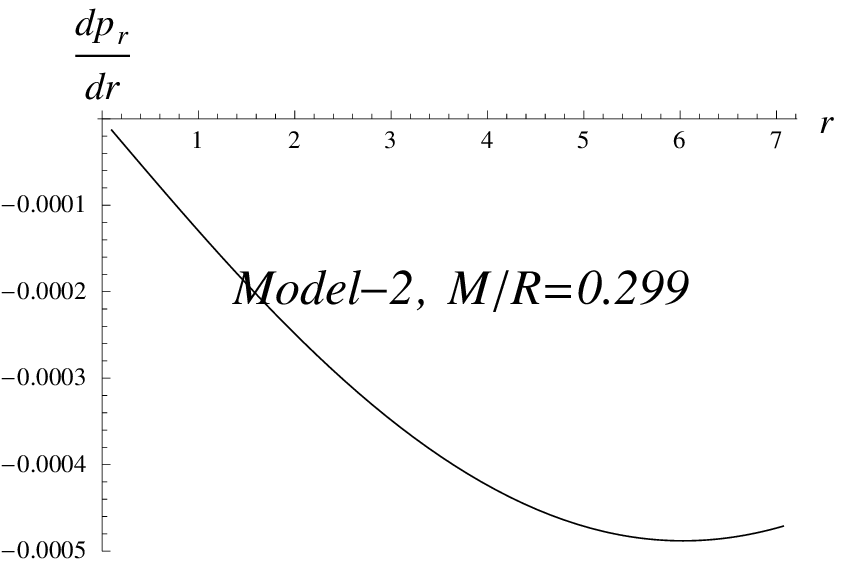, width=.36\linewidth,
height=1.4in}\epsfig{file=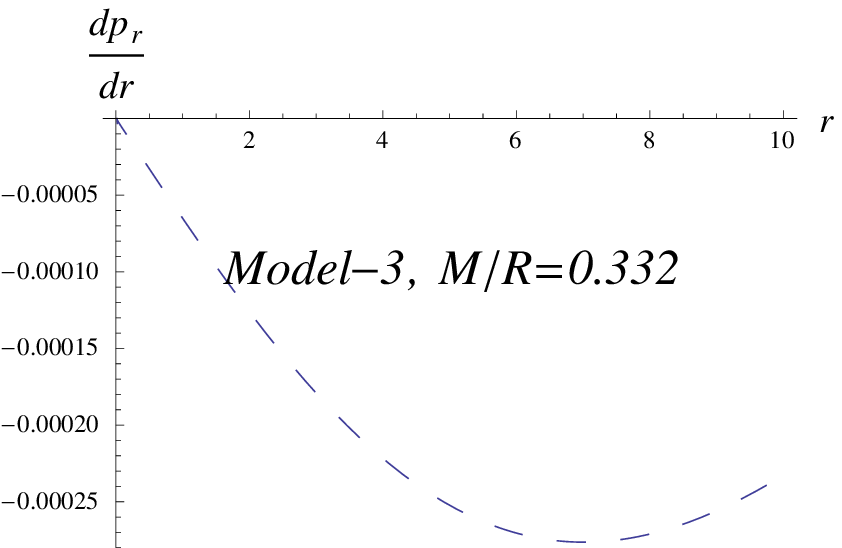, width=.34\linewidth,
height=1.4in}\caption{Evolution of $\frac{dp_r}{dr}$ with increasing radius $r$.}
\end{figure}
Using the EOS $p_r={\omega}_r \rho$ and $p_t={\omega}_t \rho$, we
obtain the following form of EOS parameters
\begin{eqnarray}\nonumber
\omega_r&=&(-e^{2Ar^2}(r^2+2\alpha(-1+\lambda))+e^{Ar^2}(4\alpha(-1
+\lambda)+ABr^6\lambda+r^2(1+8B\alpha(-1\\\nonumber&&+\lambda)-8A\alpha\lambda)
+Br^4(2+\lambda-8A\alpha\lambda+4B\alpha\lambda)-2\alpha(1+3Br^2+B^2r^4
-Ar^2(2\\\nonumber&&+Br^2))(-1+Br^2(-1+\lambda)+\lambda+B^2r^4(1+\lambda)
-Ar^2(2(1+\lambda)+Br^2(1+3\lambda)))))/\\\nonumber&&(e^{2Ar^2}(r^2+2\alpha
(-1+\lambda))+2(-1-3Br^2-B^2r^4+Ar^2(2+Br^2))\alpha(-3Br^2(1\\\nonumber&&+3\lambda)
+1-\lambda+Ar^2(-2+Br^2)(1+3\lambda)+B^2r^4(3+7\lambda))-e^(Ar^2)(4\alpha(-1
+\lambda)\\\nonumber&&+r^2(1+24B\alpha\lambda+8A\alpha(1+\lambda))+Br^6(A\lambda
+B(2+4\lambda))-r^4(A((4+8B\alpha)\lambda\\\label{13}&&+2)+B(3\lambda+4B\alpha]
(2+3\lambda))))),\\\nonumber
\omega_t&=&(-2\alpha(-1-3Br^2-B^2r^4+Ar^2(2+Br^2))(-3-B^2r^4-2\lambda
+Br^2(-1\\\nonumber&&+Ar^2)(1+2\lambda))+e^{2Ar^2}(r^2(2+\lambda)
-2\alpha(3+2\lambda))+e^{Ar^2}(4\alpha(3+2\lambda)\\\nonumber&&
+Br^6(A-B+A\lambda)-r^2(2+\lambda+4A\alpha(3+2\lambda)-4B\alpha
(5+4\lambda))+r^4(B(-1\\\nonumber&&+4B\alpha)(2+\lambda)+A(1
-8B\alpha(1+\lambda)))))/(e^{2Ar^2}(r^2+2\alpha(-1+\lambda))
+2\alpha(-1\\\nonumber&&-3Br^2-B^2r^4+Ar^2(2+Br^2))(1-\lambda
-3Br^2(1+3\lambda)+Ar^2(-2+Br^2)(1\\\nonumber&&+3\lambda)+B^2r^4
(3+7\lambda))-e^(Ar^2)(4\alpha(-1+\lambda)+r^2(1+24B\alpha\lambda]
+8A\alpha(1+\lambda))\\\nonumber&&+Br^6(A\lambda+B(2+4\lambda))
-r^4(A(2+(4+8B\alpha)\lambda)+B(3\lambda+4B\alpha(2+3\lambda))))).\\\label{14}
\end{eqnarray}
It is interesting to mention here that the EoS parameters depend on radius rather than a constant quantity as in ordinary matter distribution. This non-constant
behavior of EoS parameters is constituted by usual matter and exotic matter contributions (see Fig. \textbf{6} and \textbf{7}).
\begin{figure}
\centering \epsfig{file=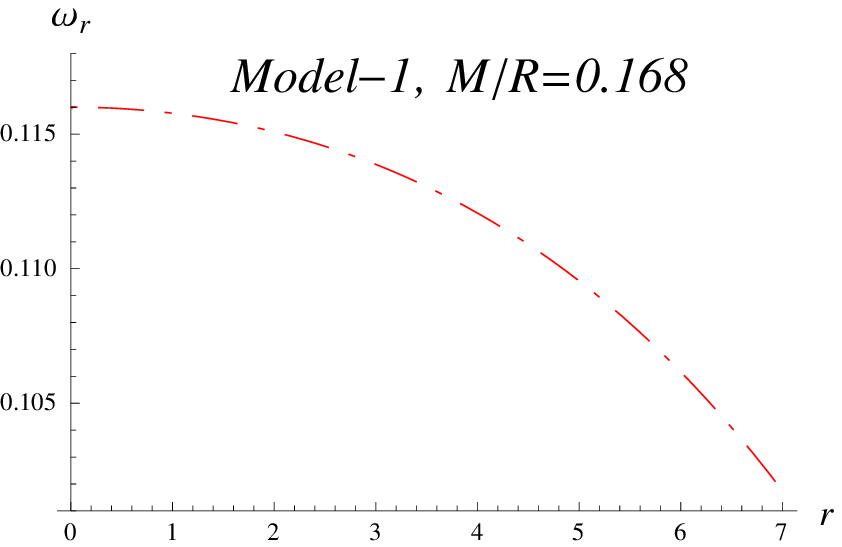, width=.36\linewidth,
height=1.4in}\epsfig{file=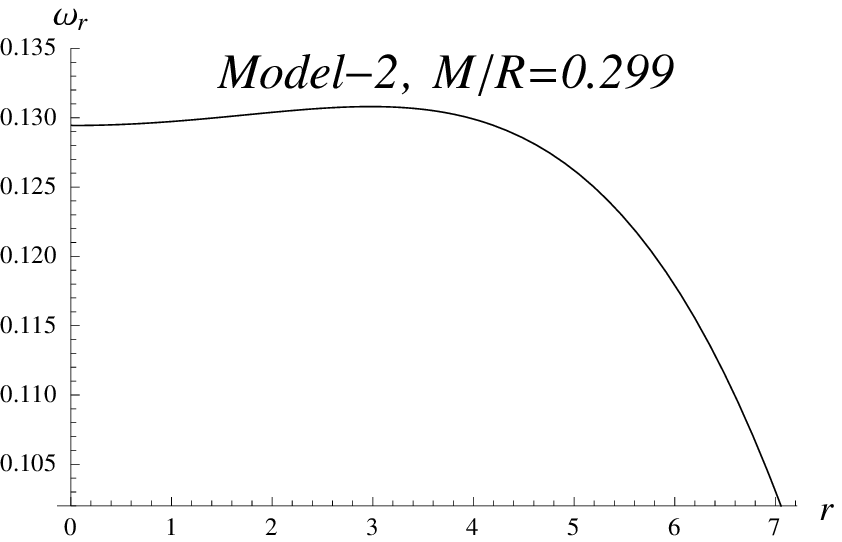, width=.34\linewidth,
height=1.4in}\epsfig{file=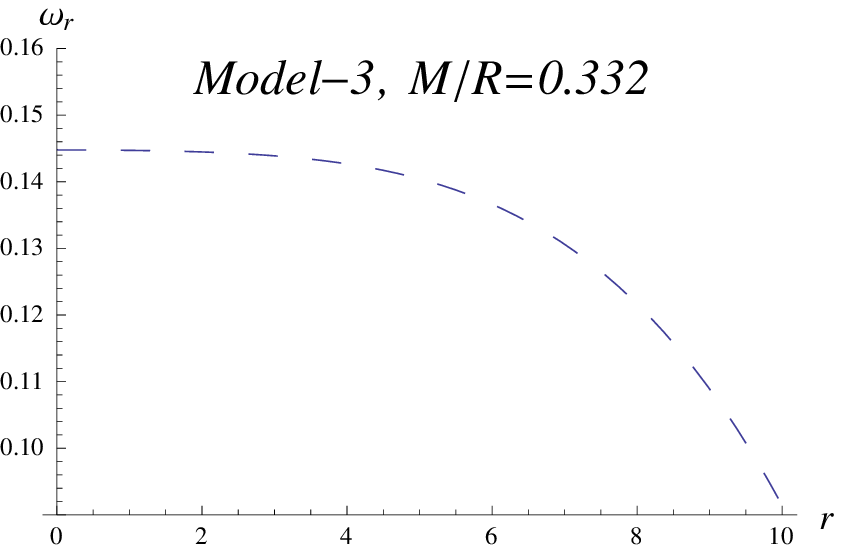, width=.36\linewidth,
height=1.4in}\caption{The evolution of radial EoS parameter across
stars.}
\end{figure}
\begin{figure}
\centering \epsfig{file=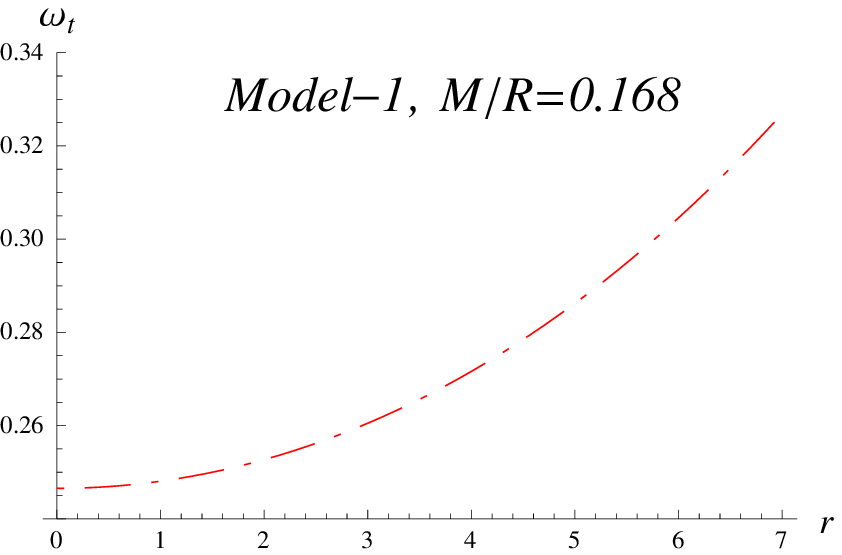, width=.34\linewidth,
height=1.4in}\epsfig{file=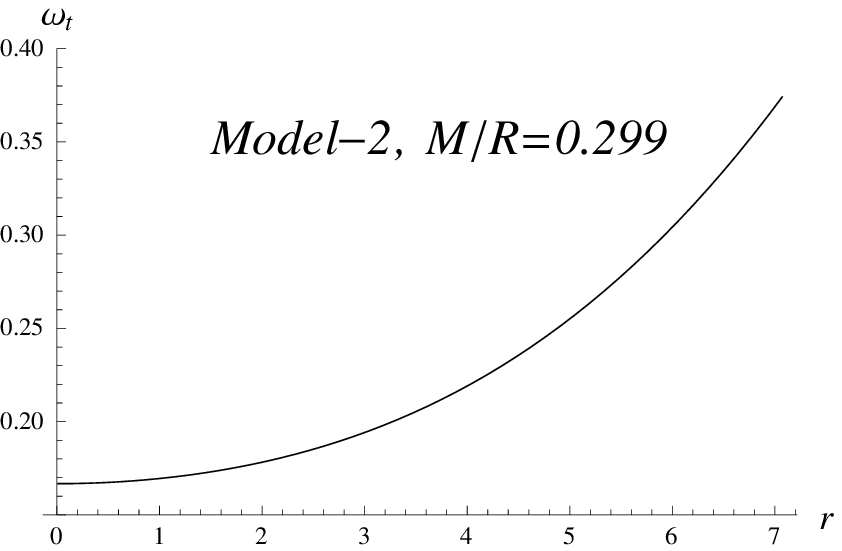, width=.36\linewidth,
height=1.4in}\epsfig{file=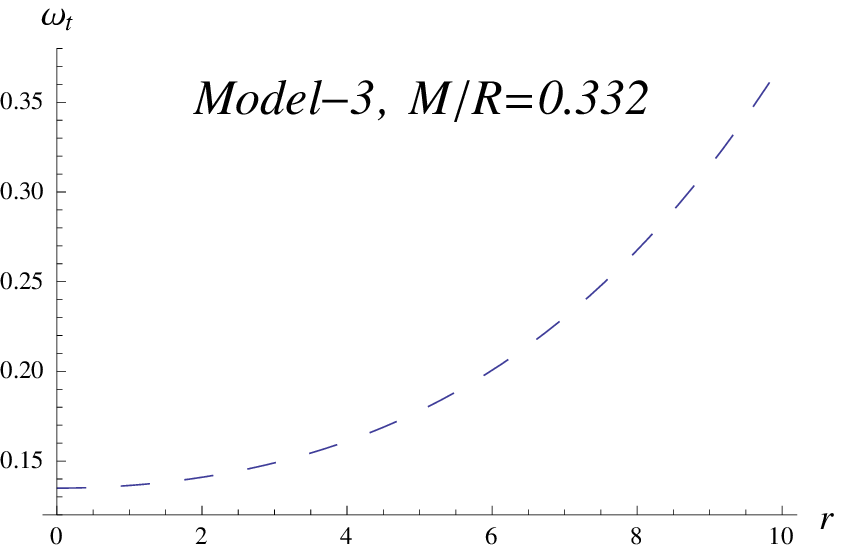, width=.34\linewidth,
height=1.4in}\caption{The evolution of tangential EoS parameter
across compact stars.}
\end{figure}\\
The measure of anisotropy $\Delta=\frac{(p_t-p_r)}{p_r}$, for
considered $f(R,T)$ model takes following form
\begin{eqnarray}\nonumber
\Delta&=&\frac{2e^{-2Ar^2}}{r^5(1+3\lambda+2\lambda^2)}\left[\left\{2\alpha(1+3Br^2+B^2r^4
-Ar^2(2+Br^2))(-3 \right.\right.\\\nonumber&&\left.\left.
-B^2r^4-2\lambda+Br^2(Ar^2-1)(1+2\lambda))+e^{2Ar^2}(r^2(2+\lambda)-2\alpha
(3+2\lambda))\right.\right.\\\nonumber&&\left.\left.
+e^{Ar^2}(4\alpha(3+2\lambda)+Br^6(A-B+A\lambda)-r^2(2+\lambda+4A\alpha(3+2\lambda)
\right.\right.\\\nonumber&&\left.\left.
-4B\alpha(5+4\lambda)))+r^4(B(4B\alpha-1)
(2+\lambda)+A(1-8B\alpha(1+\lambda)))
\right\}-\right.\\\nonumber&&\left.\left\{e^{2Ar^2}(r^2+2\alpha
(\lambda-1)+e^{Ar^2}(4\alpha(\lambda-1)+ABr^6\lambda+r^2(1+8B\alpha
(\lambda-1)\right.\right.\\\nonumber&&\left.\left.
-8A\alpha\lambda)+Br^4(2+\lambda-4\alpha\lambda(2A-B)))-
2\alpha(1+3Br^2-Ar^2(2+Br^2)
\right.\right.\\\label{16}&&\left.\left.+B^3r^4)(-1+Br^2(\lambda-1)-Ar^2(2(1
+\lambda) +Br^2(1+3\lambda))))\right\}\right].
\end{eqnarray}
Figure \textbf{8} describes the evolution of anisotropy measure,
$p_t>p_r$ i.e., $\Delta>0$ corresponds to the outward drawn
anisotropy and its directed inward when $\Delta<0$. In our model, we
find that $\Delta>0$ for the different compact stars as shown in
Figure \textbf{8}. It can be seen from Fig. \textbf{8} that
$\Delta>0$ at most of the points for compact stars Model 1 and Model
3, indicating that that a repulsive anisotropic force occurs,
allowing the construction of more massive distributions. In our
model anisotropy measure for Model 2 decreases with the increase in
radius and becomes negative beyond $r=2.8km$.

\begin{figure}
\centering \epsfig{file=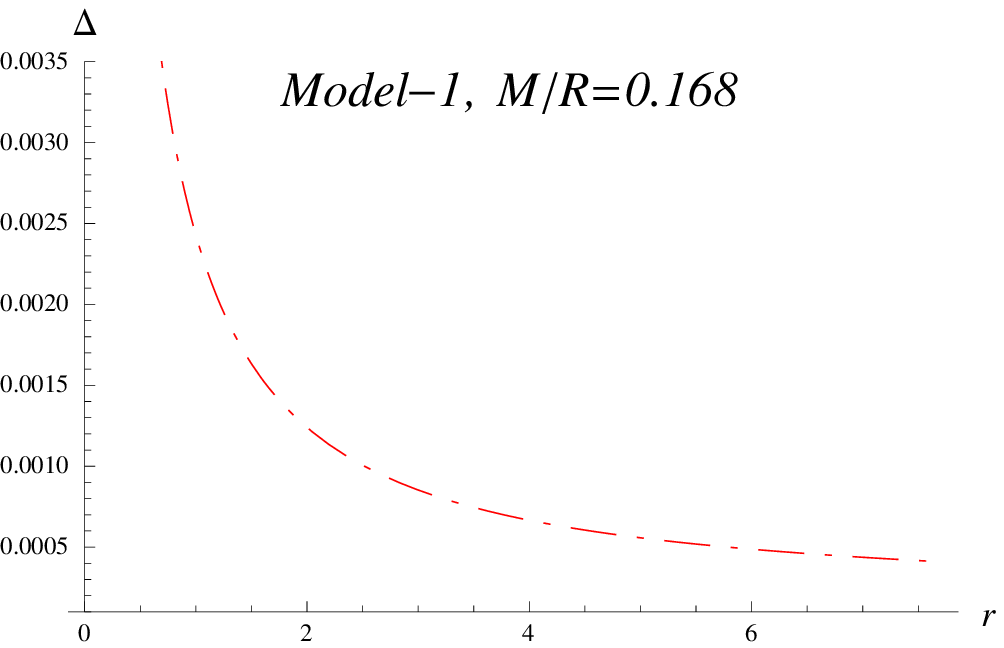, width=.34\linewidth,
height=1.4in}\epsfig{file=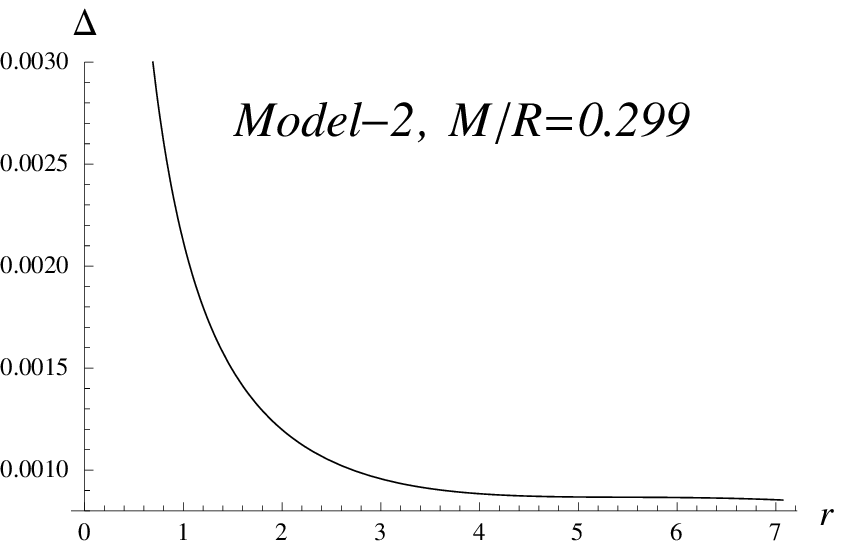, width=.36\linewidth,
height=1.4in}\epsfig{file=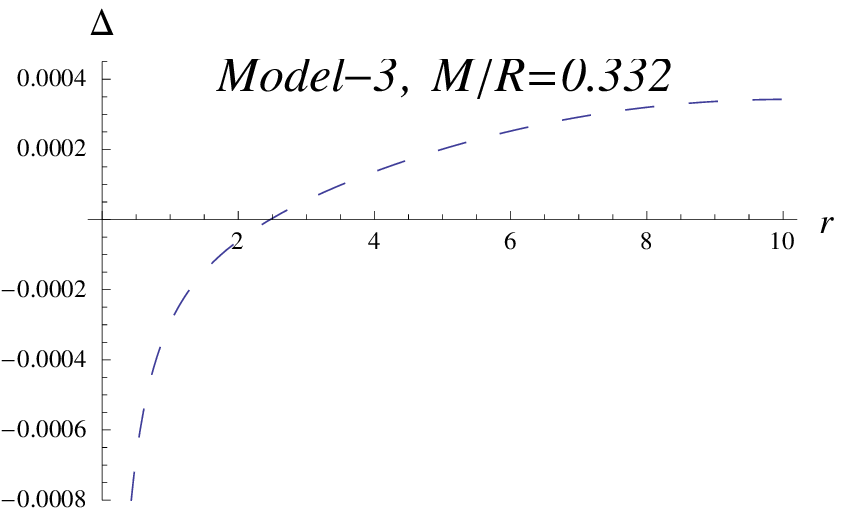, width=.34\linewidth,
height=1.4in}\caption{Anisotropy measure $\Delta$ for compact stars}
\end{figure}

\subsection{Energy Conditions}

Energy bounds are of significant importance including null energy condition (NEC), weak
energy condition (WEC), strong energy
condition (SEC) and dominant energy condition (DEC), defined as
\begin{eqnarray}\nonumber
\textbf{NEC}:\quad&&\rho+p_r\geq0, \quad \rho+p_t\geq0,\\\nonumber
\textbf{WEC}:\quad&&\rho\geq0, \quad \rho+p_r\geq0, \quad
\rho+p_t\geq0,\\\nonumber \textbf{SEC}:\quad&&\rho+p_r\geq0, \quad
\rho+p_t\geq0, \quad \rho+p_r+2p_t\geq0,\\\nonumber
\textbf{DEC}:\quad&&\rho>|p_r|, \quad \rho>|p_t|.
\end{eqnarray}
The considered anisotropic sphere satisfy the energy conditions,
exhibited graphically in Figure \textbf{9} for compact star Model 1.
\begin{figure}
\centering \epsfig{file=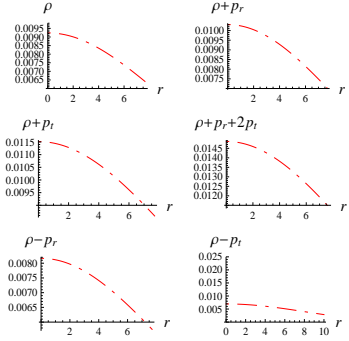, width=.34\linewidth,
height=1.8in}\caption{Energy conditions for Model 1}
\end{figure}

\subsection{Causality Conditions and Stability Analysis}

The radial and transverse sound speeds denoted by $v_{sr}$ and
$v_{st}$ should be less than speed of light i.e., $0\leq
v^2_{sr}\leq 1$, $0\leq v^2_{st}\leq 1$, where
$v^2_{sr}=\frac{dp_r}{d\rho}$ and $v_{st}=\frac{dp_t}{d\rho}$. We
plot the evolution of radial and transverse sound speeds for compact
stars and found that above mentioned conditions hold, as shown in
Fig. \textbf{10} and \textbf{11}.
\begin{figure}
\centering \epsfig{file=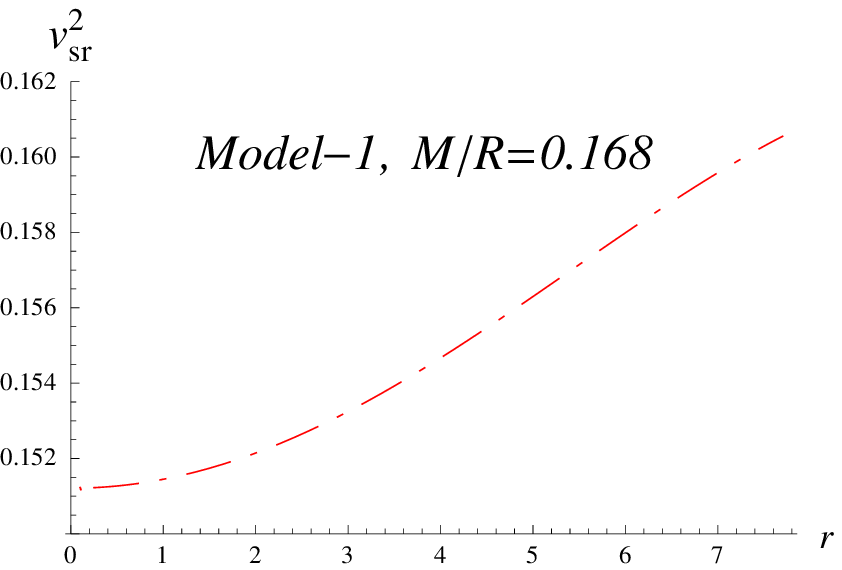, width=.34\linewidth,
height=1.4in}\epsfig{file=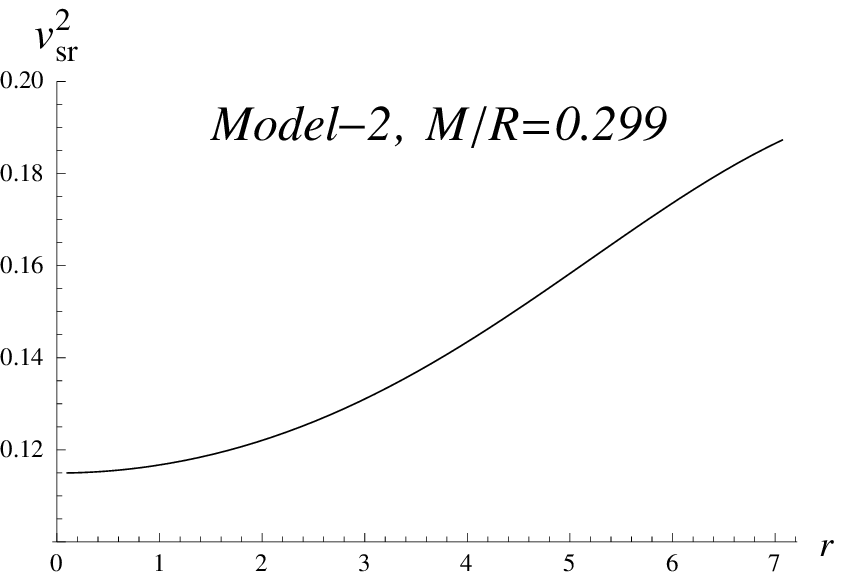, width=.36\linewidth,
height=1.4in}\epsfig{file=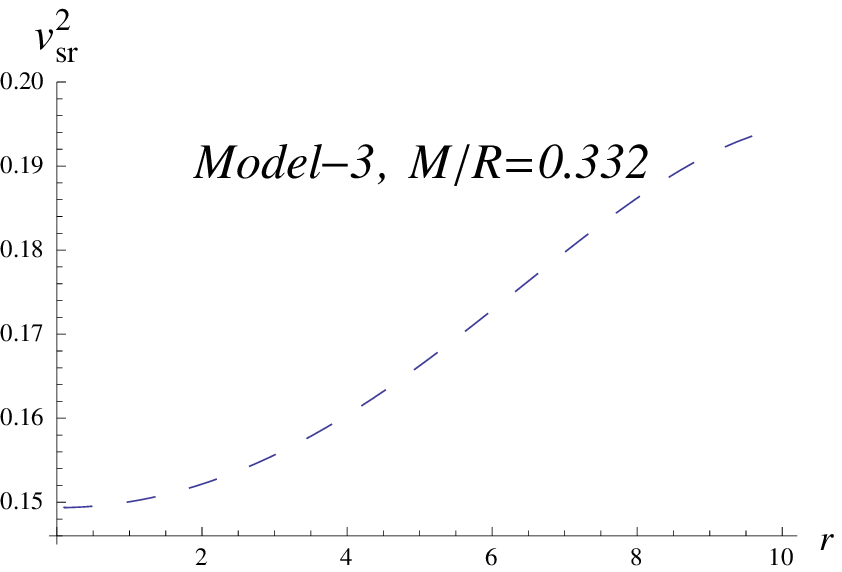, width=.34\linewidth,
height=1.4in}\caption{Plot of $v^2_{sr}$ varying with radius $r$.}
\end{figure}
\begin{figure}
\centering \epsfig{file=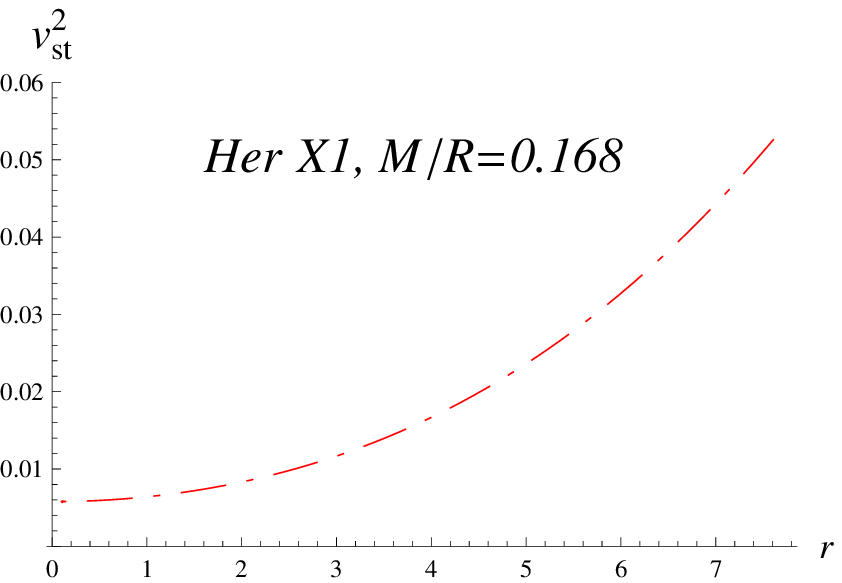, width=.34\linewidth,
height=1.4in}\epsfig{file=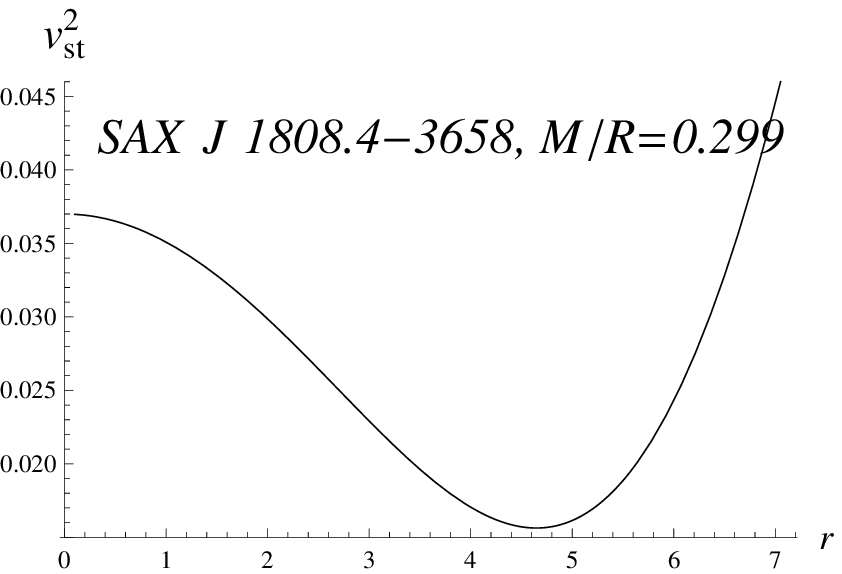, width=.36\linewidth,
height=1.4in}\epsfig{file=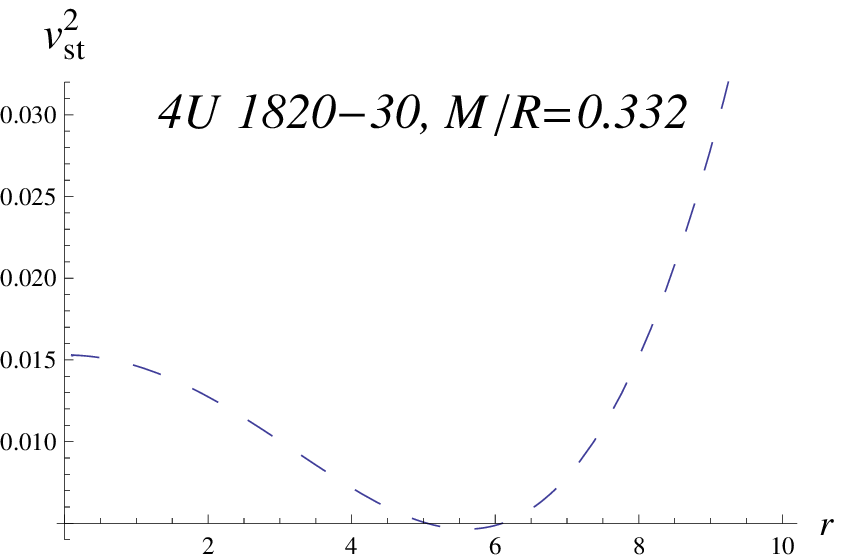, width=.34\linewidth,
height=1.4in}\caption{Plot of $v^2_{st}$ varying with radius $r$.}
\end{figure}
One can undergo with the stability analysis of compact objects
considering sound speeds (Herrera 1992, Herrera and Barreto 2013,
Herrera and Santos 1997, Herrera et al.2008). The the potentially
stable and unstable regimes can be estimated by considering the
difference of the sound propagation within the matter distribution.
The variation of $v^2_{st}-v^2_{sr}$ of different strange stars is
shown in Figure 12. Figure 12 shows that difference of the two sound
speeds, i.e., $v^2_{st}-v^2_{sr}$ retain similar sign within the
specific configuration and it satisfies the inequality
$0<|v^2_{st}-v^2_{sr}|<1$. Thus, our proposed compact stars models
are stable.

\begin{figure}
\centering \epsfig{file=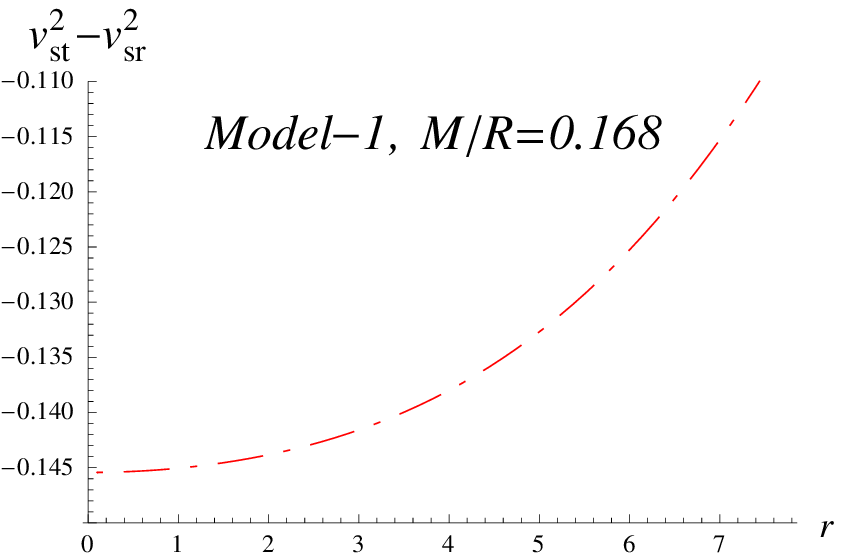, width=.34\linewidth,
height=1.4in}\epsfig{file=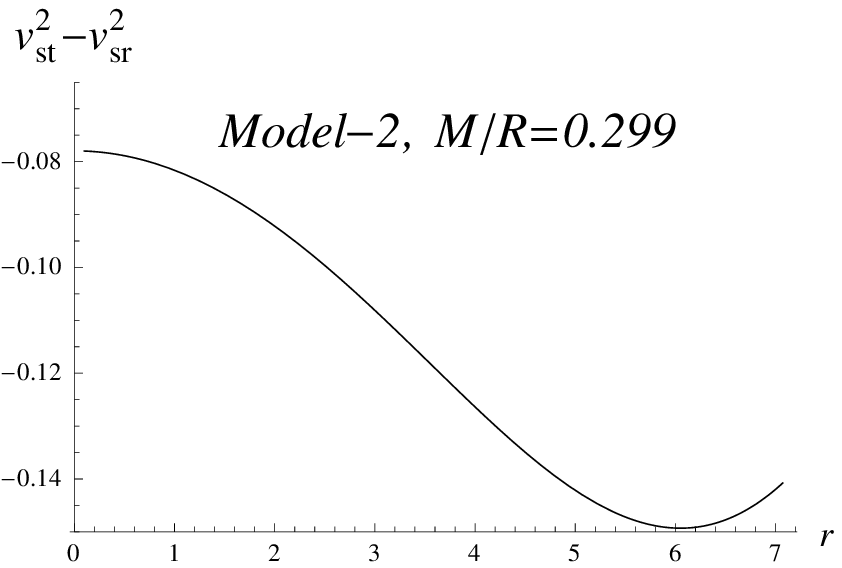, width=.36\linewidth,
height=1.4in}\epsfig{file=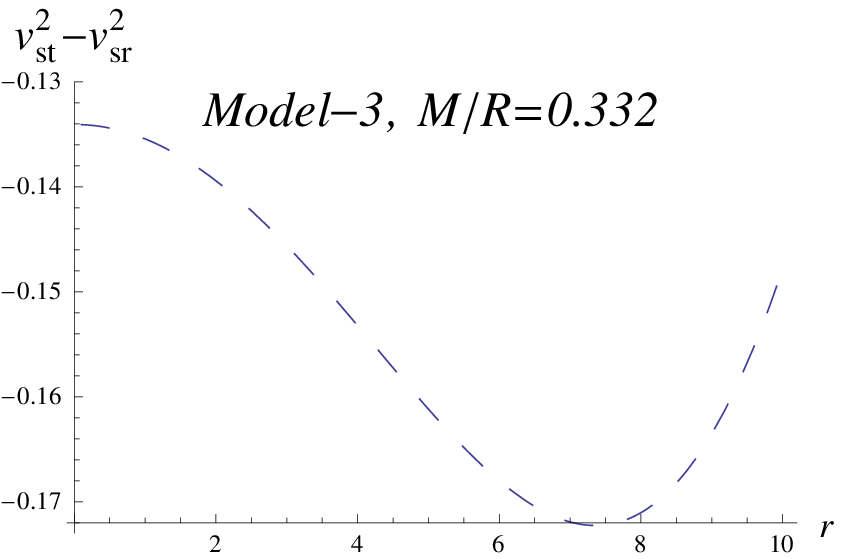, width=.34\linewidth,
height=1.4in}\caption{Plot of $v^2_{st}-v^2_{sr}$ for anisotropic
compact stars.}
\end{figure}

\section{Conclusion}

The modified theory of gravity can provide the explanation of
accelerated expansion of universe. The modified gravity theories
(such as the f(R,T) gravity) have been the center of attention for
many researchers in the recent past, because this type of theories
seems to provide a viable explanation for dark energy. The conformal
relation of $f(R,T)$ to general theory of relativity with a
self-interacting scalar field has been examined (Astashenok et al.
2014).

This paper deals with the interior solutions for anisotropic fluids,
which have been used to model compact stars in the context of
modified gravity theory $f(R,T)$. The modeling has been completed by
taking compact stars as anisotropic in $f(R,T)$ gravity. For the
exact solution of the governing differential equations, we have used
the Krori and Barua (1975) form of the metric function, i.e.,
$b=Ar^2$, $a=Br^2+C$, $A$, $B$ and $C$ are arbitrary constant that
have been calculated by using some matching conditions. The smooth
matching of interior and exterior regions of a star lead us to
express the unknown constants in terms of masses and radii of the
compact stars. Using the observed values of the masses and radii of
the compact stars, we get the values of the model constants that are
used to discuss the nature of the stars. For the calculated values
of the constant,  we found that the energy density, radial and
transverse pressure decreases for given class of compact strange
stars. For the particular choice $f(R,T)$, we have found that EOS
parameters behave like normal matter distribution. On the basis of
this fact, we may conclude that compact stars are composed of
ordinary matter and effect of $f(R,T)$ term. The regularity analysis
of the proposed model implies that density and pressure are regular
every where and attain the maximum value at the center. Thus radial
pressure$p_r$ and matter density $\rho$ have maximum values at the
center and it decreases from the center to the surface of the star,
where density becomes constant and pressure reduces to zero.

It has been found that the anisotropy will be directed outward when
$p_t>p_r$ this implies that $\Delta>0$. We have found that
$\Delta>0$ for compact stars Model 1 and Model 3, indicating that
that a repulsive anisotropic force occurs, allowing the construction
of more massive distributions. While the anisotropy measure for
Model 2 decreases with the increase in radius and becomes negative
beyond $r=2.8km$.  The the potentially stable and unstable regimes
have been estimated by considering the difference of the sound
propagation within the matter distribution. The compact stars remain
potentially stable in the regions where difference of radial and
sound speeds remain positive. It is found that
$0<|v^2_{st}-v^2_{sr}|<1$ for all three considered compact stars.
So, our considered model is potentially stable in $f(R,T)$ gravity,
as shown in Fig. \textbf{12}.

We would like to mention that TOV equations for modified theories of
gravity like $f(G)$ (Momeni and Myrzakulov 2015), non-local $f(R)$
gravity (Momeni et al. 2015) and   $f(R,T)$ (Moraes et al.2015) have
been studied analytically and numerically.

\section{Conflict of Interest}
Authors declare that there is no conflict of interest regarding the
publication of this paper.
\\
\section*{Acknowledgement}
We are thankful to the unknown referee for his/her deep study of the
paper and fruitful comments, that have improved the paper a lot.
\newpage

\newpage



\end{document}